\documentclass[twocolumn, 12pt]{openjournal}

\usepackage{xcolor}
\usepackage{textgreek}
\usepackage[utf8]{inputenc}
\usepackage[english]{babel}
\usepackage{hyperref}
\hypersetup{
    colorlinks=true,
    linkcolor=blue,
    filecolor=blue,      
    urlcolor=blue,
    citecolor=blue,
}
\usepackage{color,colortbl}
\usepackage{tensind}
\tensordelimiter{?}
\DeclareGraphicsExtensions{.bmp,.png,.jpg,.pdf}
\usepackage{verbatim}
\usepackage[normalem]{ulem}
\usepackage{orcidlink}
\usepackage{soul}

\graphicspath{ {./figs/} }

\begin{document}
\title{The effects of different cooling and heating function models on a simulated analog of NGC300}

\author{David Robinson\orcidlink{0000-0002-3751-6145}}
\email{dbrobins@umich.edu}
\affiliation{Department of Physics; University of Michigan, Ann Arbor, MI 48109, USA}
\affiliation{Leinweber Center for Theoretical Physics; University of Michigan, Ann Arbor, MI 48109, USA}

\author{Camille Avestruz\orcidlink{0000-0001-8868-0810}}
\affiliation{Department of Physics; University of Michigan, Ann Arbor, MI 48109, USA}
\affiliation{Leinweber Center for Theoretical Physics; University of Michigan, Ann Arbor, MI 48109, USA}

\author{Nickolay Y. Gnedin}
\affiliation{Particle Astrophysics Center; 
Fermi National Accelerator Laboratory;
Batavia, IL 60510, USA}
\affiliation{Kavli Institute for Cosmological Physics;
The University of Chicago;
Chicago, IL 60637, USA}
\affiliation{Department of Astronomy \& Astrophysics; 
The University of Chicago; 
Chicago, IL 60637, USA}

\author{Vadim A. Semenov\orcidlink{0000-0002-6648-7136}}
\affiliation{Center for Astrophysics $\vert$ Harvard \& Smithsonian, 60 Garden St, Cambridge, MA 02138, USA}

\begin{abstract}
 Gas cooling and heating rates are vital components of hydrodynamic simulations. However, they are computationally expensive to evaluate exactly with chemical networks or photoionization codes.  We compare two different approximation schemes for gas cooling and heating in an idealized simulation of an isolated galaxy.  One approximation is based on a polynomial interpolation of a table of Cloudy calculations, as is commonly done in galaxy formation simulations. The other approximation scheme uses machine learning for the interpolation instead on an analytic function, with improved accuracy. We compare the temperature-density phase diagrams of gas from each simulation run to assess how much the two simulation runs differ. Gas in the simulation using the machine learning approximation is systematically hotter for low-density gas with $-3 \lesssim \log{(n_b/\mathrm{cm}^{-3})} \lesssim -1$. We find a critical curve in the phase diagram where the two simulations have equal amounts of gas.  The phase diagrams differ most strongly at temperatures just above and below this critical curve. We compare \ion{C}{2} emission rates for collisions with various particles (integrated over the gas distribution function), and find slight differences between the two simulations. Future comparisons with simulations including radiative transfer will be necessary to compare observable quantities like the total \ion{C}{2} luminosity.
\end{abstract}


\maketitle

\section{Introduction}
\label{sec:intro}

Gas in the interstellar medium (ISM) and circumgalactic medium (CGM) of galaxies exists in multiple interacting phases with different temperatures and densities. These phases include a diffuse hot ionized medium \citep{spitzer56}, clouds of an atomic cold neutral medium \citep{ewen_purcell51}, a warm ionized medium surrounding the cold clouds \citep{hjellming69}, and very cold molecular clouds \citep{cheung68}. Gas cools and heats through the respective emission and absorption of radiation, providing a critical mechanism for gas to transition between these phases \citep[e.g.][]{mckee_ostriker77}.  Star formation (primarily) occurs in cold molecular clouds, and the star formation rate can depend on both the thermal state \textit{and turbulent properties} of the molecular gas (see \citet{dobbs23} and \citet{hennebelle24} for recent reviews).

Gas cooling and heating rates thereby play a vital component of galaxy evolution models, including hydrodynamic simulations. However, there are many different prescriptions in use for calculating the cooling and heating rates as functions of the gas properties and the galactic environment (see \citet{kim23} for an overview).

The effects of different cooling and heating function models on galaxy formation can be tested using galaxy formation simulations. Simulation comparisons from the Aquila and AGORA collaborations typically focus on the effects of differing stellar feedback models \citep{scannapieco12, kim14}. But, differences between varied gas thermodynamic prescriptions in simulations is relatively understudied. Previous work has explored the effect of varying the assumed initial temperature and density profiles of halo gas in semi-analytic models (SAMs), which changes how the gas subsequently cools \citep[e.g.][]{monaco14, hou18, hou19}. However, each of these works assumes a non-evolving cooling function which does not depend on the ionization state of particles in the gas. 

Some hydrodynamic simulations now incorporate \textit{non-equilibrium} chemistry, where the simulations calculate element abundances on-the-fly using a chemical network. We can also calculate gas cooling and heating rates from these non-equilibrium abundances. The assumption of gas photoionization equilibrium yields an alternate set of cooling and heating rates. \citet{richings_schaye16} compare isolated galaxy simulations run with gas cooling calculated from non-equilibrium against those run with equilibrium chemical abundances. \citet{capelo18} perform a similar comparison, but with the non-equilibrium metal abundances only used to calculate cooling in cold ($T < 10^{4} \, \mathrm{K}$) gas. Both works find that non-equilibrium cooling has little effect on the overall star formation rate of the galaxy, but can change the overall amount of molecular gas. 

In this paper, we consider the isolated galaxy simulations of an NGC300 analog presented in \citet{semenov21}. These simulations were originally used to study the spatial decorrelation between dense molecular gas and sites of recent star formation, and include subgrid models for gas turbulence and turbulence-regulated variable star formation efficiency. We use these simulations to compare two different prescriptions for approximating cooling and heating functions: the interpolation table approach from \citet{gnedin_hollon12} and the machine learning approximation from \citet{robinson24}, which yields more accurate cooling and heating function approximations at fixed metallicity.

Since the only differences between the two simulation runs are in the gas cooling and heating functions, we focus on the comparison between the resulting temperature-density phase diagrams of the simulated gas. These phase diagrams describe the gas fraction in various phases of the ISM and CGM. We also compare \ion{C}{2} emission between each simulation run. The luminosity in various ionic emission lines depend on the thermal state (i.e. temperature and density) of the gas. The $157.7 \, \mu \mathrm{m}$ fine-structure line emitted by ionized carbon (\ion{C}{2}) is a particularly important case. This line is often used to trace molecular, star-forming gas in galaxies \citep[e.g.][]{zhao24, casavecchia24}. The \ion{C}{2} fine-structure line is also a candidate for line-intensity mapping (LIM) surveys mapping the 3-D structure of the universe. LIM surveys could probe \ion{C}{2} emission across a wide range of redshifts $3 \lesssim z \lesssim 9$, stretching back into the Epoch of Reionization \citep{kovetz17}.

We explain our methodology, including the simulation code and how we analyze simulated snapshots, in Section~\ref{sec:methods}. We compare gas phase diagrams and \ion{C}{2} emission rates from our two simulation runs in Section~\ref{sec:results}, and present conclusions and further discussion in Section~\ref{sec:disc}.

\section{Methodology}
\label{sec:methods}

\subsection{Simulations} \label{method:art_ngc300}
In this work, we use the isolated galaxy simulation of \citet{semenov21}. This simulation is run using Adaptive Refinement Tree (ART), a Eulerian adaptive mesh refinement hydrodynamics code \citep{kravtsov_thesis, kravtsov02, rudd08}.

The specific simulation we use includes a sub-grid prescription for turbulence and a star formation efficiency that depends on the velocity dispersion (including both turbulent and thermal components) through the local virial parameter \citep{padoan12, semenov16, semenov21}. Different versions of the isolated galaxy simulation incorporate a constant value for star formation efficiency below a maximum virial parameter (or above a minimum density), different contributions to feedback (type II supernovae and/or pre-supernova stellar winds), and iterations with and without radiative transfer \citep{semenov21}.

The metallicity-dependent atomic gas cooling and heating functions in the simulation are evaluated using the interpolation table of \citet{gnedin_hollon12}, which interpolates between exact cooling and heating rates evaluated with the photoionization code Cloudy \citep{ferland98}. This approximation depends on the temperature $T$, baryon number density $n_b$, and metallicity $Z$ of the gas, and on rates calculated from the local radiation field: $P_\mathrm{LW}$ (the photodissociation rate of molecular hydrogen), and the photoionization rates $P_\mathrm{HI}, P_\mathrm{HeI}$, and $P_\mathrm{LW}$, in units of $[\mathrm{s}^{-1}]$ \citep{gnedin_hollon12}.

The thermal energy density $U$ of the gas evolves due to radiative processes as:
\begin{equation}
    \left. \frac{dU}{dt} \right\vert_\mathrm{rad} = n_b^2 \left[ \Gamma - \Lambda \right],
    \label{eq:chf_def}
\end{equation}
where $\Gamma$ is the heating function and $\Lambda$ is the cooling function. Note that the cooling and heating functions also include additional contributions from non-atomic components of the gas such as molecules that are not included in the interpolation table of \citet{gnedin_hollon12}, as they depend on additional properties of the gas. Examples include heating from the photodissociation of molecular hydrogen, and cooling from vibrational and rotational transitions in molecular hydrogen, which both depend on the molecular gas fraction. These contributions are treated separately in the simulation \citep{gnedin_kravtsov11}, and are not varied in this paper.

The initial conditions of the fiducial simulation are chosen to approximate observed structural properties of the NGC300 galaxy from \citet{westmeier11}. The dark matter particles are initialized with a Navarro-Frenk-White profile with mass $M_{200c} \approx 8.3 \times 10^{10} \, \mathrm{M}_\odot$ (defined as the mass contained in a sphere with average density equal to 200 times the critical density) and concentration $c_{200c} \approx 15.4$.  The stellar and gas components are initialized as (independent) exponential scale disks.  The stellar disk has mass $M = 10^9 \, M_\odot$, scale radius $1.39 \, \mathrm{kpc}$, and scale height $0.28 \, \mathrm{kpc}$. The gas disk has mass $2.29 \times 10^9 \, M_\odot$ and scale radius $3.44 \, \mathrm{kpc}$ (the scale height is determined by the ISM pressure gradient). The minimum gas cell size reached in the simulation run is $\Delta = 10 \, \mathrm{pc}$. The fiducial simulation run includes radiative transfer, all feedback processes, and a star formation efficiency depending on the local virial parameter \citep{semenov21}.

As described in Section~\ref{method:xgb_chf}, we turn off radiative transfer and use constant photoionization rates.  We also fix the gas metallicity. Following \citet{semenov21}, we initialize our simulations with a snapshot from a fiducial simulation run at $t \approx 600 \, \mathrm{Myr}$.  The fiducial simulation turns on various physical processes (such as radiative transfer, gas cooling, and star formation) in stages, allowing the galaxy to `settle' after each stage.  By $t \approx 600 \, \mathrm{Myr}$, each relevant process has been incorporated \citep{semenov21}.  After the initial snapshot, we save snapshots every $\Delta t = 1 \, \mathrm{Myr}$.  

Key parameters for our simulations are shown in Table~\ref{tab:sim_params}.  We run both simulations for the same duration in simulated cosmic time and find that the thermodynamic properties of the gas in the two simulations have sufficiently converged relative to each other after $5 \, \mathrm{Myr}$. This convergence is discussed in more detail in Appendix~\ref{ap:conv}.

\begingroup 
    \setlength{\tabcolsep}{6pt} 
    \renewcommand{\arraystretch}{1.5} 
    \begin{table}
        \centering
        \begin{tabular}{cc}
            Parameter & Value\\
            \hline
            \hline
            Radiative transfer & Off \\
            \hline
            $Q_\mathrm{LW}$ & $2 \times 10^{-11} \, \mathrm{s}^{-1}$ \\
            $Q_\mathrm{HI}$ & $2 \times 10^{-17} \, \mathrm{s}^{-1}$ \\
            $Q_\mathrm{HeI}$ & $3 \times 10^{-16} \, \mathrm{s}^{-1}$ \\
            $Q_\mathrm{CVI}$ & $9 \times 10^{-18} \, \mathrm{s}^{-1}$ \\
            $Z$ & $0.3Z_\odot$ \\
            \hline
            Timestep & $1 \, \mathrm{Myr}$\\
            \hline
            \hline
        \end{tabular}
        \caption{Important parameters for our simulation runs}
        \label{tab:sim_params}
    \end{table}
\endgroup

\subsection{Cooling and heating function models} \label{method:xgb_chf}
The \citet{semenov21} simulations calculate metallicity-dependent atomic gas cooling and heating functions using the interpolation table of \citet{gnedin_hollon12}, which incorporates a local radiation field including contributions from a synthesized stellar spectrum, quasar-like power law, and absorption by neutral hydrogen and helium. \citet{gnedin_hollon12} constructed these tables by interpolating between exact calculations of cooling and heating function from the Cloudy photoionization code \citep{ferland98}. \citet{robinson24} trained machine learning models on the same Cloudy calculations, with the same gas properties and radiation field parameters as inputs. So, we can directly replace the \citet{gnedin_hollon12} interpolation table with machine learning models from \citet{robinson24} in the simulation code. The interpolation table of \citet{gnedin_hollon12} implements a quadratic interpolation in metallicity that can lead to occasional unphysical \textit{negative} predicted cooling or heating functions. The machine learning models in of \citet{robinson24} are constructed to always predict positive cooling and heating functions.

Additionally, the interpolation table of \citet{gnedin_hollon12} is known to make `catastrophic errors' at some points in parameter space, due to interpolating a non-linear function on a grid of parameters with fixed spacing. To assess the impact of these catastrophic errors on the simulation, we replace the interpolation table with analogous machine learning models from \citet{robinson24}, constructed using the gradient-boosted tree algorithm XGBoost \citep{chen_guestrin16}. We label the simulation run using the cooling and heating function interpolation table from \citet{gnedin_hollon12} as `GH12' and the run using the analogous XGBoost models from \citet{robinson24} as `XGB'.

More specifically, we use \textit{fixed metallicity} cooling and heating function models from \citet{robinson24} with input parameters that parallel the dimensions of the \citet{gnedin_hollon12} interpolation table: temperature $T$, \textit{hydrogen} number density $n_\mathrm{H}$, and rates $Q_\mathrm{LW}$, $Q_\mathrm{HI}$, $Q_\mathrm{HeI}$, and $Q_\mathrm{CVI}$, where $Q_j = P_j/n_\mathrm{H}$. Separate models are trained to predict the cooling function and the heating function.  Both the interpolation table of \citet{gnedin_hollon12} and the machine learning models of \citet{robinson24} use quadratic interpolation in metallicity to make predictions at arbitrary metallicities, and \citet{robinson24} finds that this quadratic interpolation (with the 5 metallicity values present in the training data) is the main limitation to improving accuracy at intermediate metallicities. In order to avoid the complications of the metallicity interpolation, we fix the gas metallicity at $Z=0.3Z_\odot$ (this is one of the metallicites with exact Cloudy calculations used to train the machine learning models \citep{robinson24}).

Since the inputs (gas temperature, gas density, and 4 photoionization rates) and outputs (cooling function and heating function) at fixed metallicity are the same, we can perform a one-to-one replacement of the \citet{gnedin_hollon12} interpolation table for the machine learning models of \citet{robinson24} in the simulation code.  For this initial study, because the machine learning models of \citet{robinson24} are more computationally expensive than the \citet{gnedin_hollon12} interpolation table, we turn off radiative transfer in the ART code. So, the photoionization rates are spatially constant throughout the simulation box.  For these rates, we use ISM averages computed from the fiducial simulation run in \citet{semenov21} with radiative transfer: $Q_\mathrm{LW} = 2 \times 10^{-11} \, \mathrm{s}^{-1}$, $Q_\mathrm{HI} = 2 \times 10^{-17} \, \mathrm{s}^{-1}$, $Q_\mathrm{HeI} = 3 \times 10^{-16} \, \mathrm{s}^{-1}$, and $Q_\mathrm{CVI} = 9 \times 10^{-18} \, \mathrm{s}^{-1}$.    

The exact Cloudy calculations used to train both approximations we use are done on a grid of data points with gas temperature $10 \leq T/\mathrm{K} \leq 10^9$ and \textit{hydrogen} number density $10^{-6} \leq n_\mathrm{H}/\mathrm{cm}^{-3} \leq 10^6$ \citep{gnedin_hollon12}.  We approximate the conversion between the number densities of hydrogen and baryons as:
\begin{equation}
    n_b = \frac{1.4n_\mathrm{H}}{1 - 0.02(Z/\mathrm{Z}_\odot)}.
\end{equation}
For $Z = 0.3\mathrm{Z}_\odot$, $n_b \approx 1.41n_\mathrm{H}$. Outside of these ranges, both models are extrapolating from the Cloudy calculations. The table of \citet{gnedin_hollon12} performs a linear extrapolation from the nearest tabulated value. The machine learning approximations of \citet{robinson24} are piecewise constant, so will simply output the nearest tabulated value

To compare the approximations on the entire grid of Cloudy computations (which is used to train both models), we compute the mean squared error $\Delta = \langle (\log \mathcal{F}_\mathrm{true} - \log \mathcal{F}_\mathrm{pred})^2 \rangle$, where $\mathcal{F}$ is either the cooling function or the heating function. Note that a fraction $2.98 \times 10^{-4}$ of the grid points result in negative predicted cooling functions using the \citet{gnedin_hollon12} interpolation table. These points are removed from the mean squared error calculation.  The mean squared errors for the approximations at $Z=0.3Z_\odot$ are shown in Table~\ref{tab:chf_errors}, which shows that both models have small errors, but the machine learning models of \citet{robinson24} produce mean squared errors that are 2-3 orders of magnitude smaller at this metallicity. More details about the performance comparison can be found in \citet{robinson24}. However, this comparison is only on the data used to construct both models, so does not realistically predict their performance on new data. In the rest of the paper, we investigate the physical effect of these differences to assess how significant how they are in a simulated galaxy.

\begingroup 
    \setlength{\tabcolsep}{6pt} 
    \renewcommand{\arraystretch}{1.5} 
    \begin{table}
        \centering
        \begin{tabular}{ccc}
            & GH12 & XGB \\
            \hline 
            \hline
            Cooling & $1.69 \times 10^{-3}$ & $1.28 \times 10^{-5}$ \\
            Heating & $2.19 \times 10^{-2}$ & $3.22 \times 10^{-5}$ \\
            \hline
            \hline
        \end{tabular}
        \caption{Mean squared errors of the cooling and heating function approximations used in our simulations on the training grid of Cloudy calculations at gas metallicity $Z=0.3Z_\odot$.  The mean squared errors for the XGBoost models from \citet{robinson24} are 2-3 orders of magnitude smaller than those for the \citet{gnedin_hollon12} interpolation table.}
        \label{tab:chf_errors}
    \end{table}
\endgroup

Note that, while we can directly replace calls of the interpolation table of \citet{gnedin_hollon12} with evaluations of machine learning models from \citet{robinson24} in the ART code, this has a major computational cost. Evaluating each timestep of the \citet{semenov21} isolated galaxy simulation (see Section~\ref{method:art_ngc300} for details) is nearly 20 times slower with the machine learning models. While it is possible that this slowdown could be reduced with careful optimization of the simulation run with machine learning models, a one-to-one replacement for simulations run over a longer timescale or a cosmological volume is simply not practical. For these applications, an interpolation table that can be evaluated quickly \textit{and an understanding of the effects of the errors made by this interpolation} are crucial. But, it is much easier to incorporate additional photoionization rates as inputs in the machine learning framework of \citet{robinson24} (which includes models using as many as 22 photoionization rate features) than to add an additional dimension to an interpolation table.

\subsection{CII luminosity} \label{method:cii_luminosity}
From the simulation data, we can compute quantities that depend on the density and temperature of the gas.  As a representative example, here we consider \ion{C}{2} emission.  When collisionally excited, \ion{C}{2} ions can emit a $157.7 \, \mu\mathrm{m}$ photon.  Setting this photon energy equal to $kT$ yields a temperature of $91.2 \, \mathrm{K}$ \citep{draine_ism_book}.  

To compare the two simulation runs, we compute the ratio of CII emission due to various excitation channels, $j$:
\begin{equation}
    r_j = \frac{\int f_\mathrm{XGB}(n_b, T)R_j(T) \, dT}{\int f_\mathrm{GH12}(n_b, T)R_j(T) \, dT}.
    \label{eq:cii_emission_ratios}
\end{equation}
where $f(n_b, T)$ is the density-temperature distribution function (the phase diagram, normalized by the total gas mass) and $R_j(T)$ is a temperature-dependent rate factor that encodes the likelihood of a collision inducing the emission of a $157.7 \, \mu\mathrm{m}$ photon, with units $[\mathrm{erg} \, \mathrm{cm}^3 \, \mathrm{s}^{-1}]$. The excitation channels $j$ include free electrons, elemental hydrogen and helium, molecular hydrogen $\mathrm{H}_2$, and CMB photons \citep{draine_ism_book}. Note that the only difference between the numerator and denominator of Equation~(\ref{eq:cii_emission_ratios}) is the simulation run used to determine the distribution function, while $R_j(T)$ remains the same.  

Since $R_j(T)$ appears inside a temperature integral in both in the numerator and denominator of Equation~(\ref{eq:cii_emission_ratios}), we only need the temperature dependence of $R_j(T)$ and can neglect multiplicative constants. For $j=\mathrm{H}_2$ (i.e. for collisions with molecular hydrogen), we separate the two spins of molecular hydrogen, and have \citep{draine_ism_book}:
\begin{equation}
    R_{\mathrm{H}_2 \, \mathrm{para}}(T) \propto \left(\frac{T}{100 \, \mathrm{K}}\right)^{0.124 - 0.018 \ln{(T/(100 \, \mathrm{K}))}},
    \label{eq:h2_para_rate}
\end{equation}
and
\begin{equation}
    R_{\mathrm{H}_2 \, \mathrm{ortho}}(T) \propto \left(\frac{T}{100 \, \mathrm{K}}\right)^{0.095 + 0.023 \ln{(T/(100 \, \mathrm{K}))}}.
    \label{eq:h2_ortho_rate}
\end{equation}

For collisions with atomic hydrogen, we use: \citep{barinovs05}
\begin{equation}
    R_\mathrm{H}(T) \propto e^{-91.2 \, \mathrm{K}/T}\left(16 + 0.344 \sqrt{\frac{T}{1 \, \mathrm{K}}} - \frac{47.7 \, \mathrm{K}}{T}\right).
\end{equation}
The rate for helium $R_\mathrm{He}(T)=0.38R_\mathrm{H}(T)$ \citep{draine_ism_book}, and so has identical dependence on the gas temperature.

Collisions with free electrons result in the rate:\citep{tayal09, draine_ism_book}
\begin{equation}
    R_e(T) \propto e^{-91.2 \, \mathrm{K}/T} \sqrt{\frac{1 \, \mathrm{K}}{T}}.
\end{equation}

Finally, \ion{C}{2} emission can be excited by CMB photons at a rate which depends on the CMB temperature $T_\mathrm{CMB}$ \citep{draine_ism_book}:
\begin{equation}
    R_\mathrm{CMB}(T) \propto e^{-91.2 \, \mathrm{K}/T_\mathrm{CMB}}.
    \label{eq:cmb_rate}
\end{equation}
Since the CMB temperature is independent of the local gas temperature, $R_\mathrm{CMB}(T) = {\rm const}$.

\section{Results} \label{sec:results}
Here, we present results comparing our two simulation runs, using the cooling and heating function interpolation table from \citet[][labelled `GH12']{gnedin_hollon12} and the analogous XGBoost models from \citet[][labelled `XGB']{robinson24}.  We show results from our final snapshot, $5 \, \mathrm{Myr}$ after our starting snapshot. This is sufficiently long for the new steady-state temperature-density distribution to settle (see Appendix~\ref{ap:conv}).

\subsection{Phase diagrams and residuals} \label{res:phase_diagrams}
We begin by calculating residuals between the temperature-density phase diagrams of gas in each simulation. Without radiative transfer, simulation snapshots include the gas mass density $\rho$, thermal energy density $u_\mathrm{th}$, and molecular weight $\mu$. From these, we calculate the baryon number density $n_b$ and temperature $T$ as follows
\begin{equation}
    n_b = \frac{\rho}{m_p},
    \label{eq:number_density}
\end{equation}
\begin{equation}
    T = \frac{2}{3} \frac{\mu}{kn_b}u_\mathrm{th},
\end{equation}
where $m_p$ is the mass of a proton and $k$ is Boltzmann's constant.  

We select 175 logarithmically spaced bins in temperature with $10 < T/\mathrm{K} < 10^{8}$ and 200 logarithmically spaced bins in baryon number density with $10^{-7} < n_b/\mathrm{cm}^{-3} < 10^{3}$ to include nearly all of the gas in the simulation box. 

\begin{figure}
    \centering   \includegraphics[width=\columnwidth]{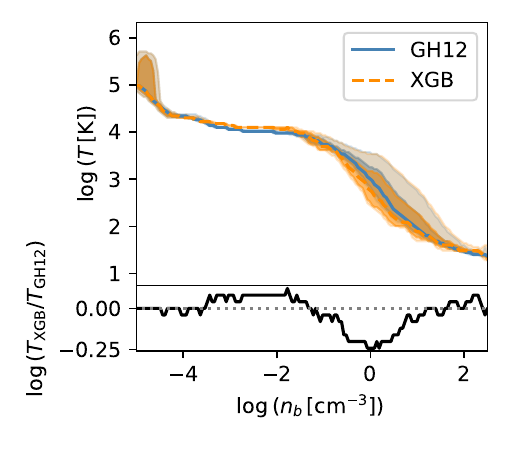}
    \caption{Upper panel: Median (curves), 25th-75th percentile (darker bands), and 10th-90th percentile (lighter bands) temperatures as a function of gas density bin for GH12 (blue, solid curve) and XGB (orange, dashed curve) runs. Bottom panel: ratio of median temperature for the XGB run to the GH12 run as a function of density bin.}
    \label{fig:temp_quantiles}
\end{figure}

In each density bin, we calculate the median gas temperature, as well as the 10th, 25th, 75th, and 90th percentiles to show the spread. These temperature quantiles are shown for both simulation runs as a function of gas density in the upper panel of Fig.~\ref{fig:temp_quantiles}. The ratio of the median temperatures for the two runs is shown in the bottom panel of Fig.~\ref{fig:temp_quantiles}. The distributions are generally similar, but the temperature for the XGB run is systematically higher for low gas densities $-3 \lesssim \log{(n_b/\mathrm{cm}^{-3})} \lesssim -1$. The bottom panel of Fig.~\ref{fig:temp_quantiles} shows that the median temperature for the XGB run is generally \textit{lower} than for the GH12 run at higher gas densities $-1 \lesssim \log{(n_b/\mathrm{cm}^{-3})} \lesssim 1$. However, at these densities, the 25th-75th percentile spread is larger than the difference between median temperatures, so the two simulations do not have systematically different temperatures here.

\begin{figure}
    \centering
    \includegraphics[width = \columnwidth]{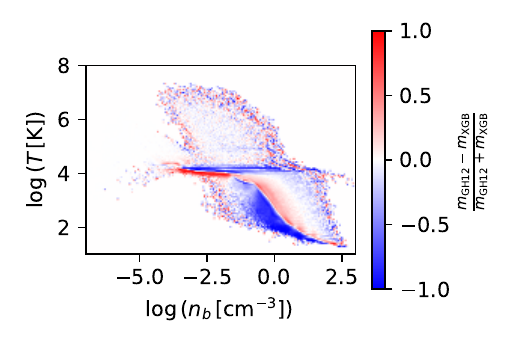}
    \caption{Residual of gas mass between runs using the interpolation table of GH12 and XGBoost cooling and heating functions after $5 \, \mathrm{Myr}$.  There is a `critical curve' where the residual is equal to $0$.}
    \label{fig:res_5_Myr}
\end{figure}

To further examine the difference in gas thermodynamics between the two simulations, we evaluate the residual in each temperature-density bin:
\begin{equation}
    \Delta = \frac{m_\mathrm{GH12} - m_\mathrm{XGB}}{m_\mathrm{GH12} + m_\mathrm{XGB}},
    \label{eq:phase_residuals}
\end{equation}
where $m_j$ is the gas mass in a temperature-density bin for simulation $j$.  The sign of the residual $\Delta$ indicates which simulation run has a higher gas mass in a given bin.  These residuals are shown for our final snapshot in Fig.~\ref{fig:res_5_Myr}.

Fig.~\ref{fig:res_5_Myr} shows that there is a `critical curve' where $\Delta = 0$, indicating that the two simulation runs have identical gas mass in the bins along this curve.  The critical curve runs from temperatures of $T \sim 10^{4} \, \mathrm{K}$ at a density of $n_b \sim 10^{-4} \, \mathrm{cm}^{-3}$ down to $T \sim 10 \, \mathrm{K}$ at $n_b \sim 10^2 \, \mathrm{cm}^{-3}$, and has a non-trivial shape. Just above this curve, the GH12 run has higher gas masses ($\Delta > 0$), while just below it, the XGB run has higher gas masses ($\Delta < 0$).  There is also a band at $T \sim 10^{4} \, \mathrm{K}$ across several orders of magnitude in density where the XGB run has higher gas masses ($\Delta < 0$).  This band and the residuals on either side of the critical curve are the most prominent structures in Fig.~\ref{fig:res_5_Myr}.  At temperatures $T \gtrsim 10^4 \, \mathrm{K}$, there is noise with no clear structure in phase space.

\begin{figure}
    \centering
    \includegraphics[width = \columnwidth]{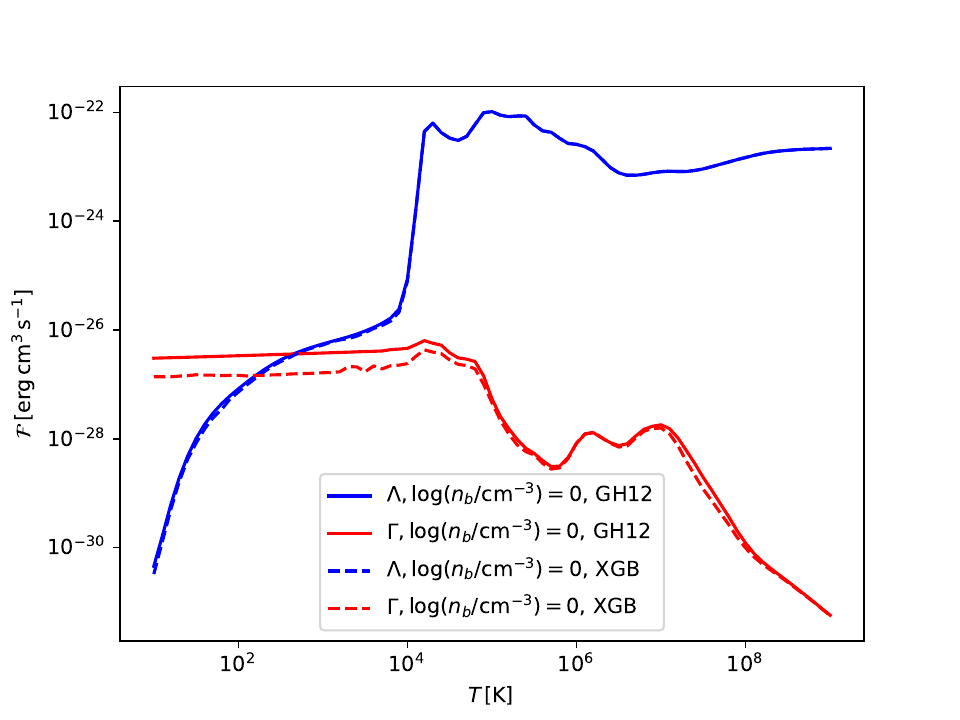}
    \caption{Predicted cooling (blue curves) and heating functions (red curves) for the GH12 interpolation table \citep[][solid curves]{gnedin_hollon12} and XGBoost machine learning models \citep[][dashed curves]{robinson24} at $n_b = 1 \, \mathrm{cm}^{-3}$. The cooling and heating curves for a given model intersect at the equilibrium temperature.}
    \label{fig:fixed_n_b_cfs}
\end{figure}

To better understand the critical curve in Fig.~\ref{fig:res_5_Myr}, we examine the temperature dependence of cooling and heating functions at fixed density $n_b = 1 \, \mathrm{cm}^{-3}$ calculated using the same approach as in the GH12 and XGB simulation runs. This is shown in Fig.~\ref{fig:fixed_n_b_cfs}. Note that the gas reaches thermal equilibrium at a temperature $T_\mathrm{equib}$ when $\Gamma(T_\mathrm{equib}) = \Lambda(T_\mathrm{equib})$ (i.e., when the cooling and heating functions are equal) if no external work is done on it. Thus, the curves for a given model in Fig.~\ref{fig:fixed_n_b_cfs} intersect at the equilibrium temperature for that model.

Fig.~\ref{fig:fixed_n_b_cfs} shows that XGBoost predicts a lower equilibrium temperature $T_\mathrm{eq}$ than does GH12 for $\log(n_b/\mathrm{cm}^{-3})=0$.  At some value between the two predicted equilibrium temperatures, the two simulation runs will have equal gas masses (the values are consistent with the `critical curve' in Fig.~\ref{fig:res_5_Myr}). Just below this critical temperature, $\Gamma > \Lambda$ for GH12, so the gas in the GH12 simulation heats up and gets above the critical temperature. For the XGB simulation, $\Lambda > \Gamma$, so the gas can cool down to its equilibrium temperature. So, the XGBoost simulation should have higher gas mass just below the critical curve, as we see in Fig.~\ref{fig:res_5_Myr}. A similar argument predicts that the GH12 simulation should have higher gas mass just above the critical curve.

\subsection{Effects on CII Luminosity} \label{res:emiss}
\begin{figure}
    \centering
    \includegraphics[width=\columnwidth]{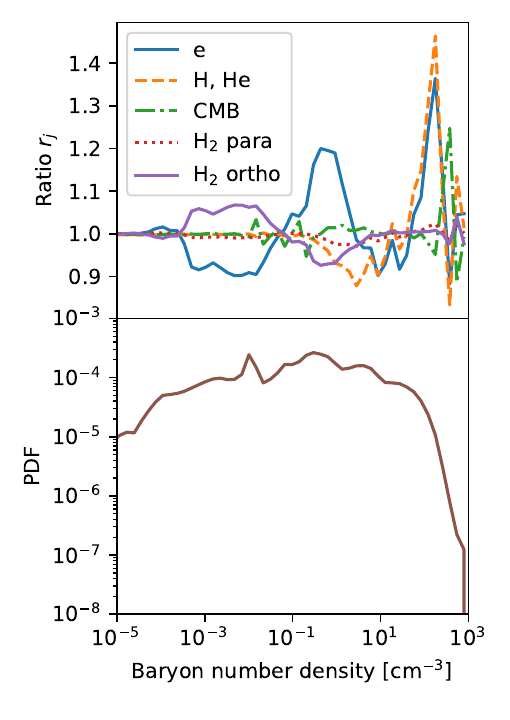}
    \caption{The \ion{C}{2} emission ratio $r_j$, as defined in Equation~(\ref{eq:cii_emission_ratios}) as a function of baryon number density $n_b$ (upper panel) exicted by interactions with electrons (solid blue), atomic hydrogen or helium (dashed orange), CMB photons (dash-dotted green), and molecular hydrogen with para (dashed red) and ortho (solid purple) spins. For comparison, the bottom panel shows the overall baryon number density profile for the GH12 simulation run (solid brown).}
    \label{fig:cii_ratios}
\end{figure}

In addition to seeing differences in the temperature-density phase diagram of simulated gas in Section~\ref{res:phase_diagrams}, we would also like to understand how these differences affect observable properties of the gas. To investigate this, we plot the ratio $r_j$ defined in Equation~(\ref{eq:cii_emission_ratios}) versus $n_b$ in the upper panel of Figure~\ref{fig:cii_ratios} for the various processes contributing to the \ion{C}{2} emission described in Section~\ref{method:cii_luminosity}. In the lower panel, we show the overall density profile for the GH12 simulation run (the profile for the XGB run is nearly identical) to compare the size of the difference in \ion{C}{2} emission with where most of the gas lies in phase space.

As shown in Figure~\ref{fig:cii_ratios}, the density profile is nearly flat between $-4 \lesssim \log(n_b/\mathrm{cm}^{-3}) \lesssim 2$, and sharply decreases for $\log(n_b/\mathrm{cm}^{-3}) \gtrsim 2$.  The ratios $r_j$ are always of order $1$ for all \ion{C}{2} emission processes.  The largest values $r_j \approx 1.4$ are reached for collisions with electrons and atomic hydrogen or helium at densities $\log(n_b/\mathrm{cm}^{-3}) \approx 2$, where the gas mass has already begun to decrease from its value at lower densities. At the lower densities where the gas mass is at is plateau for both simulation runs, $r_j$ differs noticeably from $1$ for collisions with electrons (as low as $0.9$ and as large as $1.2$), ortho H$_2$ (as low as $0.95$ and as large as $1.05$), and atomic hydrogen and helium (as low as $0.9$).  At a density of $n_b \sim 10^2 \, \mathrm{cm}^{-3}$, all processes shown have $r_j \gtrsim 1$. So, we would expect the actual \ion{C}{2} luminosity to be different between XGB and GH12 runs for gas at these densities (and, in particular, for the XGB gas to have a higher \ion{C}{2} luminosity). At other densities, some processes have $r_j \gtrsim 1$ while others have $r_j \lesssim 1$, so it is not clear whether the two simulations would have systematically different \ion{C}{2} luminosities.

\section{Summary and discussion}
\label{sec:disc}
In this paper, we implement the machine learning approximation scheme for radiation-field-dependent cooling and heating functions (at a fixed metallicity) of \citep{robinson24} into the ART hydrodynamic simulation code, as an alternative to the interpolation table of \citet{gnedin_hollon12}.  We compare runs of an isolated galaxy simulation of an NGC300 analog using the two different approximations, without radiative transfer and at fixed gas metallicity. We start the simulations from a fiducial simulation snapshot after the galaxy has `settled' once all relevant physical processes have been added to the simulation.  We compare the gas thermodynamics in the two simulations after $5 \, \mathrm{Myr}$ using their temperature-density phase diagrams. Our main conclusions are:

\begin{itemize}
    \item The gas in the XGB simulation is systematically hotter for low-density gas with $-3 \lesssim \log{(n_b/\mathrm{cm}^{-3})} \lesssim -1$ (see Fig.~\ref{fig:temp_quantiles}).
    \item There is a `critical curve' where the two simulation runs have identical gas masses. The largest differences in simulated gas mass occur at temperatures just above and below this critical curve (see Fig.~\ref{fig:res_5_Myr}).
    \item At a given density, the critical curve lies at a temperature between the equilibrium temperatures corresponding to the predicted GH12 and XGB cooling and heating functions at that density (see Fig.~\ref{fig:fixed_n_b_cfs})
    \item The differences in gas thermodynamics result in small, but not necessarily negligible (10-20\%) differences in the integrated \ion{C}{2} rate for some emission processes (see Fig.~\ref{fig:cii_ratios}).
\end{itemize}

The net cooling function (the cooling function minus the heating function) determines the temperature of a gas cloud.  The thermal properties of the gas are one of the factors that determine its velocity dispersion.  In the simulations that we use, the local star formation efficiency is calculated using the local velocity dispersion \citep{semenov16, semenov21}. So, the differences between the thermal properties of the gas in the two simulations can propagate to differences in the local star formation rate and stellar feedback. To explore these effects in a more realistic setting, we would need simulations including radiative transfer and varying metallicity,  over a long enough timescale to look for effects on the star formation history of the galaxy.  With radiative transfer, it would also be possible to directly compute \ion{C}{2} luminosities for the two simulated galaxies. Here, we take a step towards this goal by investigating the simplified case of gas thermodynamics in a `settled' isolated galaxy simulation with spatially constant metallicity and photoionization rates. This case is interesting because it allows us to explore the direct impact of changes in the gas cooling and heating functions. In a more realistic simulation, this would be difficult to disentangle from the effects of other processes that depend on gas thermodynamics (such as star formation and feedback). Even in this simplified case, we see differences in the gas thermodynamics and \ion{C}{2} emission efficiencies for several emission processes. 

In a more realistic case with varying gas metallicity across the simulated galaxy, we would expect the difference between the XGB and GH12 runs to be smaller than seen here. This is because the difference in performance between the two approximations is smaller across a sample of points with arbitrary metallicity \citep{robinson24}. However, future machine learning cooling and heating approximations (with access to more training data in metallicity) will likely be able to improve the performance at arbitrary metallicity.

As described in Section~\ref{method:xgb_chf}, the direct implementation of the \citet{robinson24} XGBoost models we used in the simulation code resulted in timesteps that were an order of magnitude slower than for simulation using the interpolation table of \citet{gnedin_hollon12}.  So, utilizing the \citet{robinson24} models in simulations of more massive galaxies or cosmological volumes is computationally infeasible. However, the machine learning approach of \citet{robinson24} much more easily accommodates additional radiation field dimensions. Adding an additional input to an interpolation table of \citet{gnedin_hollon12} would require increasing the dimension of the table by one, increasing the size of the table that must be stored in memory. Every node of the regression trees used in XGBoost models only considers the value of one input feature \citep{chen_guestrin16}. The tree depth and number of trees used in the trained models of \citet{robinson24} do not increase with the number of inputs. This makes it feasible to train models with large numbers of photoionization rate features. So, in cases where more features describing the radiation field are needed, machine learning models could be less computationally expensive than an interpolation table. The machine learning setup of \citet{robinson24} also allows for the calculation of `feature importance' values describing how much each input affects model predictions. \citet{robinson24} uses these feature importances to construct additional machine learning models using different sets of 4 photoionization rates than the set used by the \citet{gnedin_hollon12} interpolation table. This approach can inform the identification of new sets of radiation field properties for constructing new, more accurate cooling and heating function interpolation tables in various regimes. 

With these computational constraints, we only considered an isolated galaxy simulation in this paper.  However, the effects of different cooling and heating function models could depend on galaxy mass, galaxy environment, and redshift.  To explore these effects, future work could extend the comparison of simulations with different cooling and heating function models to both a suite of multiple isolated galaxies, and a cosmological volume.\\

\section*{Acknowledgments}
The authors would like to thank Raziq Noorali for his assistance calculating \ion{C}{2} emission rates due to various processes from simulation data.

DR and CA acknowledge support from the Leinweber Foundation.  CA acknowledges support from DOE grant DE-SC009193. This manuscript has been co-authored by Fermi Research Alliance, LLC under Contract No. DE-AC02-07CH11359 with the U.S. Department of Energy, Office of Science, Office of High Energy Physics. Support for VS was provided by Harvard University through the Institute for Theory and Computation Fellowship. This research was also supported in part through computational resources and services provided by Advanced Research Computing (ARC), a division of Information and Technology Services (ITS) at the University of Michigan, Ann Arbor, in particular the Great Lakes cluster and the U-M Research Computing Package.

This work utilizes several Python packages, including XGBoost \citep{chen_guestrin16}, yt \citep{yt11}, and numpy \citep{numpy}. The code used to analyze the simulation data and produce the plots can be found at \url{https://github.com/davidbrobins/ngc300_analysis}.

\bibliographystyle{apsrev4-1}

\bibliography{oja_template}

\begin{appendix}

\section{Convergence of the gas phase diagram}
\label{ap:conv}
As discussed in Section~\ref{method:art_ngc300}, we start our simulation runs from fiducial snapshots after the simulated galaxy has settled. So, we need only be concerned with the convergence of the gas thermodynamics due to the different cooling and heating function models. To assess this, we compute a residual between $\Delta$ (see Equation~\ref{eq:phase_residuals}) across two snapshots $a$ and $b$ (note that the two simulation runs have snapshots at identical times):
\begin{equation}
    \delta_{a,b} = \frac{\Delta_b - \Delta_a}{\left| \Delta_a + \Delta_b \right|}.
    \label{eq:squared_resids}
\end{equation}
Since the residuals at the two timesteps $\Delta_a$ and $\Delta_b$ can be positive or negative (see Fig.~\ref{fig:res_5_Myr}), the absolute value in the denominator is crucial to ensure that $-1 \leq \delta_{a,b} \leq 1$.  For convergence, the phase space map of $\delta_{a,b}$ should have no structure. That is, the value of $\delta_{a,b}$ in nearby bins should be as uncorrelated as possible.

\begin{figure*}
    \centering   
    \includegraphics[width = 0.49\textwidth]{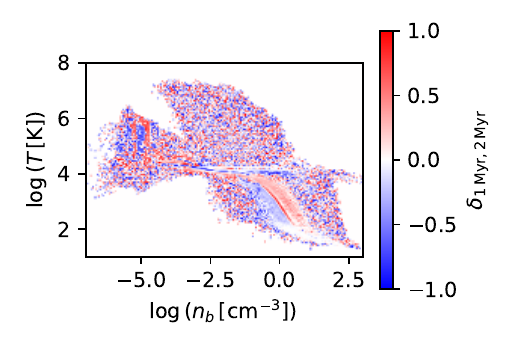} 
    \includegraphics[width = 0.49\textwidth]{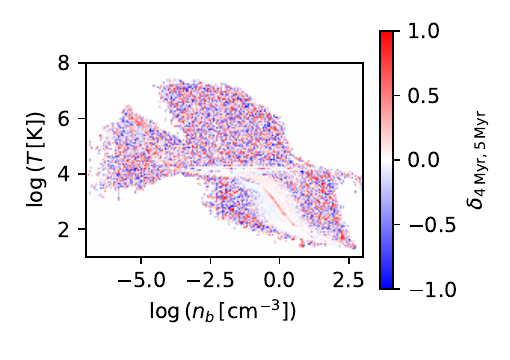}
    \caption{Residual between residuals $\delta_{1 \, \mathrm{Myr}, 2 \, \mathrm{Myr}}$ (left, see Equation~(\ref{eq:squared_resids})) and $\delta_{4 \, \mathrm{Myr}, 5 \, \mathrm{Myr}}$ (right).}
    \label{fig:squared_res}
\end{figure*}

To demonstrate the convergence of gas thermodynamics by $5 \, \mathrm{Myr}$, we show $\delta_{1 \, \mathrm{Myr}, 2 \, \mathrm{Myr}}$ at left and $\delta_{4 \, \mathrm{Myr}, 5 \, \mathrm{Myr}}$ at right in Fig.~\ref{fig:squared_res}. Going from $1$ to $2 \, \mathrm{Myr}$ in the left panel of Fig.~\ref{fig:squared_res}, clear structure can be seen in vertical stripes between $-6 \lesssim \log(n_b/\mathrm{cm}^{-3}) \lesssim -4$ and $4 \lesssim \log(T/\mathrm{K}) \lesssim 6$, as well as a similar structure to the critical curve seen in Fig.~\ref{fig:res_5_Myr}.  By the timestep from $4$ to $5 \, \mathrm{Myr}$ in the right panel of Fig.~\ref{fig:squared_res}, the vertical stripes are no longer visible, and the critical curve structure is both much thinner and peaks at lower values of $|\delta|$.  The remaining regions in Fig.~\ref{fig:squared_res} show noise with no clear phase space structure.  So, we can say that the gas thermodynamics of our simulation runs have sufficiently converged after $5 \, \mathrm{Myr}$, and all of the analysis in Section~\ref{sec:results} use the final $5 \, \mathrm{Myr}$ timestep.

\end{appendix}

\end{document}